# Ta-based *S-I-S* junctions for integrated superconducting electronics

Northrop Grumman Quantum Computing Team*

Northrop Grumman Mission Systems, Linthicum Heights, MD, 21090, USA

*See acknowledgements for author list.

(Dated: September 9, 2026)

**ABSTRACT**

Josephson junctions built around metal-oxides other than $AlO_x$ or $NbO_x$ have gained a resurgence of interest given needs to address loss-mechanisms in superconducting qubits. In this work we show that *S-I-S* Josephson junctions with a $TaO_x$ tunnel barrier and α-Ta electrodes can be grown using seed layer engineering and fabricated to satisfy a wide range of critical current densities, thus demonstrating that they are an interesting candidate to develop and evaluate for application in classical and quantum superconducting electronics. Further, we show that these $TaO_x$ junctions can be fully realized with industry compatible methods at wafer scale using a combination of cryogenic test and room temperature wafer probing techniques.

## I. Introduction

Josephson junctions provide the required nonlinearity for superconducting technologies, such as qubits, superconducting digital logic, amplifiers, and SQUIDs. Aluminum-oxide ($AlO_x$) and niobium-oxide ($NbO_x$) based junctions have been the overwhelmingly used material stack for these devices. However, $NbO_x$ junctions exhibit suboptimal sub-gap resistance, providing an undesigned and undesirable damping and power consumption to the devices. Additionally, while superconducting qubits have demonstrated coherence times up to a millisecond [1] using $AlO_x$ junctions, two-level systems (TLS) in the amorphous junction barrier result in strongly-coupled interactions with the qubit mode [2, 3], resulting in rare, but catastrophic, degradation of device performance. These considerations motivate exploring novel junction stacks. Using Ta and Ta-based compounds to study the impact on Josephson junction performance is of particular interest in both the electrodes and oxide barrier, given recent literature incorporating Ta as the trace material for superconducting microwave resonators [4, 5, 6, 7, 8, 9, 10, 11, 12] or as the capacitor pads in transmon qubits [4, 1, 13, 14]. Further, from a processing point of view, Ta may prove advantageous because it does not have the strict temperature budgets associated with Al-based junctions grown on Si that are due to the tendency of Al pitting into Si and excessive hillock formation [15]; additionally, Ta is robust to aggressive chemical cleans. Here, we consider two tantalum-based junctions as alternative options for Josephson junctions.

In this work we report on the reliable creation of fully Ta-based *S-I-S* junctions at the 150 mm wafer scale. We combine insights from previous literature that have studied aspects of this materials problem in isolation to experimentally realize *S-I-S* junctions in which the base electrode (BE) and top electrode (TE) both primarily consist of α-Ta [16, 17, 18, 19, 20], and the insulating tunnel barrier separating the superconductors is $TaO_x$ [21, 22, 23, 24, 25]. We achieve the α-Ta electrodes through seed layer engineering [26, 27, 28, 29, 30], and a controlled, tunable $TaO_x$ process reliant on plasma oxidation techniques [19]. We show that these stacks exhibit electrical tunneling behavior characteristic of *S-I-S* Josephson junctions. The methods used to demonstrate these fully Ta-based junctions rely on Si substrates, sputter deposition, and standard lithographic processes—all of which are compatible with modern industrial foundries. Finally, by developing and characterizing a plasma oxidation technique that allows the thickness and morphology of the $TaO_x$ layer in the stack to be reproducibly and precisely controlled, we illustrate the feasibility of tuning the critical current density ($J_c$) of these Ta-based junctions to achieve critical current ($I_c$) values as low as 40 nA for a 0.4 $\mu m^2$ junction, which meets the requirements for integration into state-of-the-art qubit architectures.

## II. Results and Discussion

### i. Thin film growth and characterization of α-Ta/$TaO_x$/α-Ta films

We first demonstrate that our α-Ta film growth is consistent with the rest of the literature in phase formation, electrical properties, and superconducting $T_c$ as measured by resistivity and Meissner screening. All Ta-based films discussed in this work were sputter deposited onto 150 mm (100)-Si substrates at room temperature; additional growth details are discussed in the Methods section. We show that our Ta films grown directly on Si crystallize in the β-phase with X-ray diffraction (XRD) (see Figure 1a, black curve), as expected [31, 32, 33, 34]. It has been widely reported that α-Ta can be preferentially nucleated on Si using metallic seed layers, which we repeat in this work using Nb or $TaN_x$ seed layers. In order to seed α-Ta, we use Nb seed layers because of its well-reported ability to promote the α-phase [26, 27] and its processing compatibility with the foundry facility used in this work. Additionally, we selected $TaN_x$ as an alternative seed layer to Nb because of its demonstrated performance as a diffusion barrier in copper metallization on Si [35, 36], and because the existence of several stable niobium-oxide states and prevalence of niobium-silicides lead to inconsistent metal-substrate and metal-air surfaces. The $TaN_x$ composition selected in this work was chosen based upon the maximum observed $T_c$ for single-layer 300 Å-thick films, in conjunction with its reliability in nucleating α-Ta when used as a seed layer [28, 29, 30]. Figure 1a shows 2Θ-ω XRD scans for 1 kÅ-thick Ta layers sputtered on top of 50 Å-thick seed layers of Nb and $TaN_x$, respectively. In both cases, we observe characteristic BCC α-Ta Bragg peaks that correspond to a *HH0* crystallographic direction preferentially oriented near the out-of-plane direction.

We measured the electrical resistivity ($\rho$) and magnetic moment ($m$) as a function of temperature for Ta films grown on both the Nb and $TaN_x$ seed layers characterized by XRD in Figure 1 — (a) Coupled 2θ-ω XRD scans of a Ta film grown directly on Si (black) and a seed layer of either Nb (blue) or $TaN_x$ (red). (b) Resistivity as a function of temperature for the same Nb- and $TaN_x$-seeded Ta films, with the inset showing a zoom in near the superconducting transition. (c) Magnetic moment as a function of temperature for both Nb- and $TaN_x$-seeded Ta films. Samples were zero field cooled to 2 K, then warmed in a small perpendicular applied field to 6 K, then cooled again to 2 K with the field still applied. No measurable superconducting transition was observed for Ta grown directly on the Si substrate down to 2 K, and so is not included in panels (b) and (c).Figure 1(a). Note that we do not show $\rho(T)$ or $m(T)$ data for the Ta sample grown directly on Si because $\rho(T)$ remains nearly constant at 180 μΩ-cm from 300 K to 2 K, and no superconducting transition was observed above 2 K, as expected for a majority β-phase Ta film [37]. By contrast, $\rho(T)$ measurements for the Nb- and $TaN_x$-seeded Ta films both exhibit the behavior of a conventional metal; the measured residual resistivity ratios are ρ(295 K)/ρ(5 K) = 25 μΩ-cm/8.3 μΩ-cm = 3.0 (Nb) and 28 μΩ-cm/12 μΩ-cm = 2.3 ($TaN_x$), respectively. These films transition into zero-resistance superconducting states at temperatures of $T_c$ = 4.30 K and 4.05 K for Nb-seeded and $TaN_x$-seeded, respectively; both $T_c$s are comparable to the known $T_c$ of 4.48 K observed for bulk α-Ta [38, 39], with the small suppressions of $T_c$ typically ascribed to the high levels of quenched structural disorder that are frequently present in thin films [11, 40]. The Meissner expulsion measurements, shown in Figure 1(c), exhibit superconducting onset temperatures of 4.25 K and 4.00 K that are consistent with the resistivity data, again demonstrating the Nb-seeded Ta film showing a slightly higher $T_c$ than $TaN_x$-seeded Ta. However, Meissner effect measurements provided a more comprehensive picture of the entire superconducting thin film; transitions are broadened and extend down to approximately 2.6 K in both samples; this likely indicates that these Ta films are not purely α-phase, but rather include a minor volume fraction of crystallites with lower $T_c$, which affects the magnetic penetration. Taken together, we find the crystallographic, resistivity, and Meissner measurements are consistent in showing that Nb- and $TaN_x$-seed layers are suitable for obtaining primarily α-Ta films with reasonably high $T_c$, which can be accessed without requiring a dilution refrigerator. We note that while the proposed application space for Ta-based Josephson junctions does not require such high temperature operation, the ability to screen devices for characteristic behavior without requiring millikelvin temperatures is beneficial in and of itself.

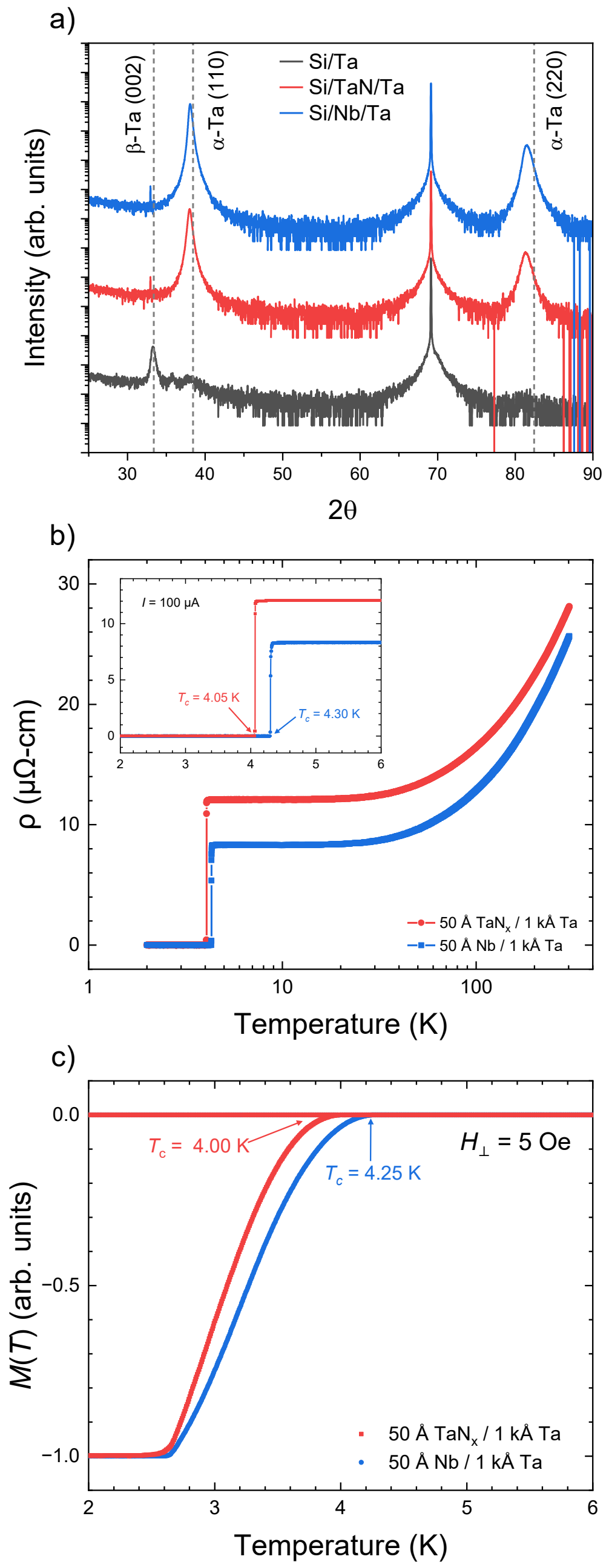

Figure 1 — (a) Coupled 2θ-ω XRD scans of a Ta film grown directly on Si (black) and a seed layer of either Nb (blue) or $TaN_x$ (red). (b) Resistivity as a function of temperature for the same Nb- and $TaN_x$-seeded Ta films, with the inset showing a zoom in near the superconducting transition. (c) Magnetic moment as a function of temperature for both Nb- and $TaN_x$-seeded Ta films. Samples were zero field cooled to 2 K, then warmed in a small perpendicular applied field to 6 K, then cooled again to 2 K with the field still applied. No measurable superconducting transition was observed for Ta grown directly on the Si substrate down to 2 K, and so is not included in panels (b) and (c).

Using the scheme discussed above for nucleating α-Ta films, we now move to growth and characterization of *S-I-S* film stacks. Our *S-I-S* samples take the general form: Si substrate / seed layer / α-Ta / $TaO_x$ / β-Ta / re-seed layer / α-Ta, where the seed (and re-seed) layers are Nb or $TaN_x$. We show that the re-seed layers for the top Ta electrodes are necessary since predominantly (00*L*)-oriented β-Ta top electrodes are observed when grown directly after oxidizing the base α-Ta electrodes (Figure 2, teal curves). When inserted into the stack above the amorphous $TaO_x$ layer, 30 Å Nb or $TaN_x$ layers again stabilize the α-Ta phase deposited above, in agreement with previous studies that employed Nb (also W & Al) re-seed layers to realize α-Ta-$TaO_x$-α-Ta (α-Ta-$AlO_x$-α-Ta) *S-I-S* stacks, respectively [16, 17, 18, 19]. From a crystal structure perspective, this re-seeding effect is illustrated in Figure 2 by the enhanced integrated intensities of the α-Ta (110) Bragg peaks for the green *S-I-S* samples relative to the base-electrode-only counterparts [40]. Finally, we note that the remaining 30 Å-thick β-Ta layer between the $TaO_x$ and re-seed layer is employed as a preventative measure against oxygen diffusion from $TaO_x$ into the seed layer for the top α-Ta film.

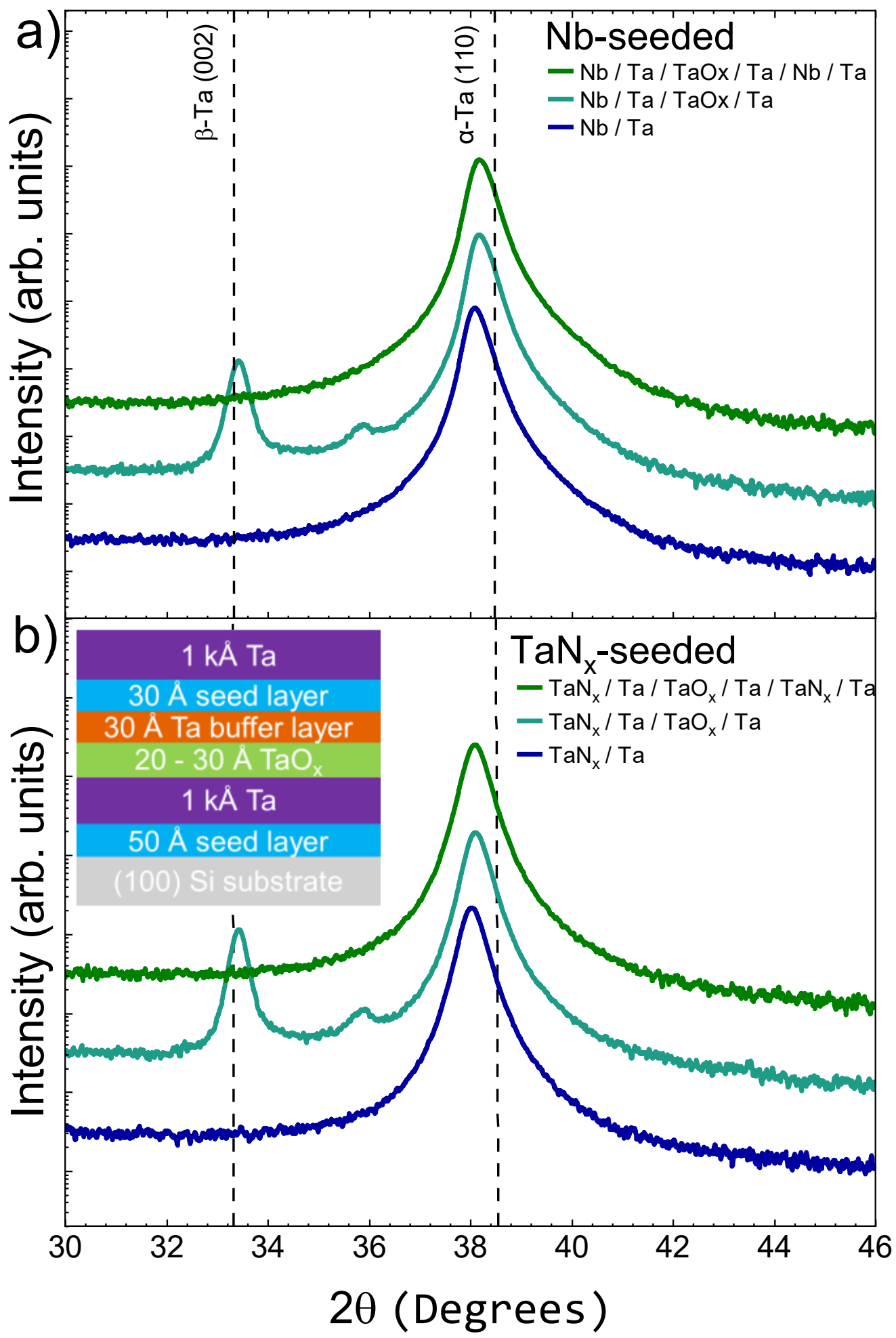

Figure 2 – 2Θ-ω XRD scans for stacks using (a) Nb- and (b) $TaN_x$-seed films. Scattering intensities are plotted with a logarithmic scale and are vertically offset for readability. The equivalent base-electrodes of seeded α-Ta are shown as references. The inset shows a cartoon schematic of the completed layer structure with base and top electrode films. Ta top electrodes deposited directly on the $TaO_x$ barrier (no re-seed layer, teal data) show characteristic β-Ta peaks centered near 2Θ = 33.4° and 35.8° [33, 34]. Re-seeding the top electrode is shown in green (which results in primarily α-Ta throughout the entire stack.

The $TaO_x$ tunnel barriers in our *S-I-S* stacks were grown by oxidizing the base electrodes with an Ar + $O_2$ plasma oxidation process. The mean thicknesses of the $TaO_x$ layers were measured using X-ray reflectometry (XRR) on simplified structures and span between 10 Å and 30 Å, which correspond to plasma oxidation durations of 5 seconds to 25 seconds (XRR data and fits are discussed in the supplementary materials [40]). We note that we have also investigated additional oxidation pathways for creating a $TaO_x$ barrier including: RF-sputtering from a $Ta_2O_5$ target, radical shower oxidation, and natural oxidation. Of these approaches, the plasma oxidation process consistently proved to be the most repeatable and highest quality for our junctions. The details of these alternative approaches are outside the scope of this work, and agree with the *S-I-S* materials literature that suggests techniques which incorporate high-energy ion bombardment, such as glow discharge oxidation [21, 22, 24, 25] and plasma oxidation [19], are the most reliable methods for growing $TaO_x$ tunnel barriers with abrupt interfaces. Further, unlike a natural oxidation process, $TaO_x$ grown by plasma oxidation is not limited by the Mott-Cabrera properties of oxidizing Ta, which self-limits at approximately 2 to 3 nm at room temperature [19, 41]. A full summary of sample stack structures with nominal and measured thicknesses are discussed in Supplementary Table 1 [40].

Based upon the discussion above, we arrive at the complete nominal junction structures in this work as: Si/Nb(50 Å)/Ta(1 kÅ)/$TaO_x$(*t*)/Ta(30 Å)/ Nb(30 Å)/Ta(1 kÅ) or Si/$TaN_x$(50 Å)/Ta(1 kÅ)/ $TaO_x$(*t*)/Ta(30 Å)/$TaN_x$(30 Å)/Ta(1 kÅ), where *t* is the $TaO_x$ thickness and ranges from approximately 10 Å to 30 Å, and is characterized using scattering and current density relations, and is discussed in later sections and the supplementary materials. The patterning and device electrical properties will be discussed in the following sections.

### ii. *S-I-S* Josephson junction devices

We patterned our junctions in circular pillars, with areas between 0.4 and 3.1 $\mu m^2$, using deep ultra-violet (DUV, 248 nm) photolithography and reactive ion etching. Contacts to the base and top Ta electrodes are made by routing a Nb wiring layer over tetraethyl orthosilicate-based $SiO_x$ (TEOS) with etched vias extending down to the Ta electrodes. A cartoon schematic and tilted scanning electron microscopy (SEM) image of a fabricated device are shown in Figure 3. Higher magnification images of the $TaO_x$ barrier, diffusion barrier, and re-seed layers using scanning transmission electron microscopy (STEM) in high-angle annular dark field (HAADF) mode, are shown in Figure 4, which show consistent oxide barriers for both Nb- and $TaN_x$-seeded devices. Elemental micrograph mappings for Ta, O, N, and Nb were collected using energy dispersive X-ray spectrometry (EDS) and electron energy loss spectrometry (EELS) and are presented in Figure 4(c) and Figure 4(f); line scans to show the elemental profiles as a function of depth from the top to base electrode are shown in Supplementary Figure 3 [40].

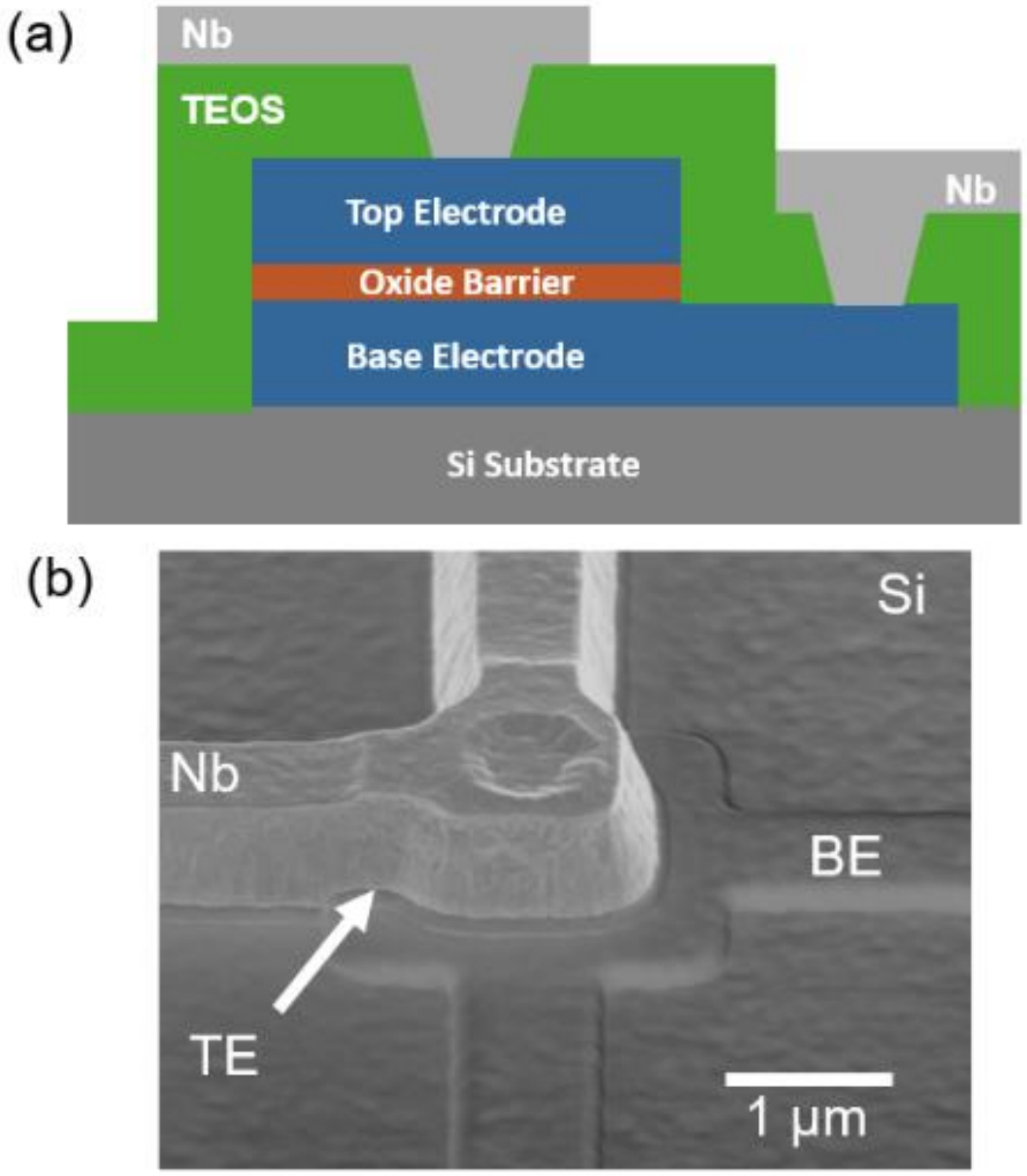

Figure 3 – (a) Cartoon schematic of fabricated *S-I-S* junctions. (b) Tilted scanning electron micrograph showing a 2 μm$^2$ *S-I-S* junction post-fabrication. The field of view shows the Si substrate, base and top Ta electrodes, and the $TaO_x$ barrier can be faintly seen by the surface topology as the surfaces are coated in TEOS. A higher resolution scanning transmission electron micrograph of the $TaO_x$ barriers, from a separate device, are shown in Figure 4.

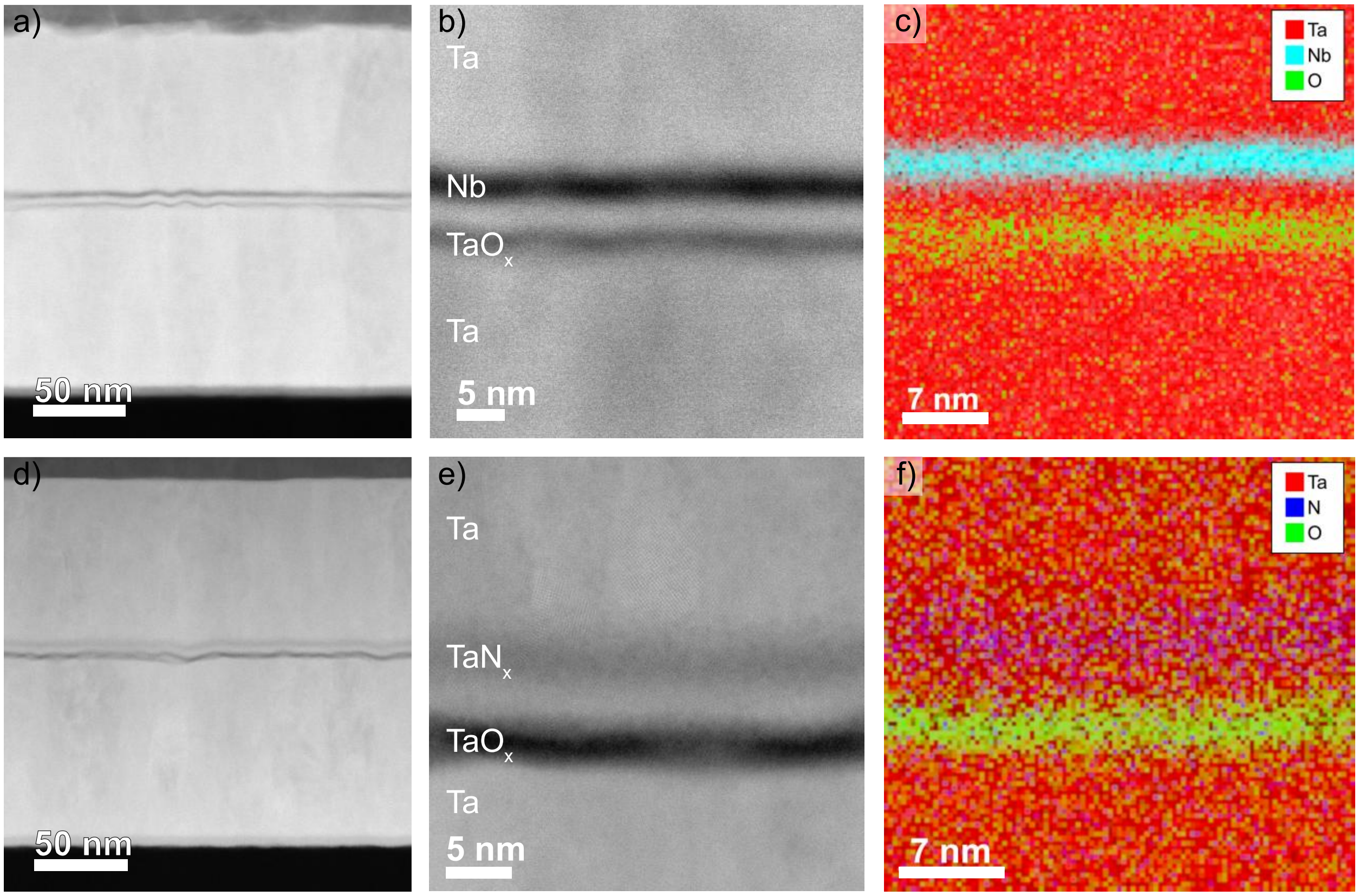

Figure 4 — STEM images of the patterned junction devices. The top row is for the Nb seeded junctions, and the bottom row shows the $TaN_x$ seeded. (a,d) HAADF STEM cross-section showing the entire stack structure and (b,e) a zoom in of the $TaO_x$ barrier and seed layer for the top electrode. (c,f) Elemental maps of the oxide barrier, and re-seed layers for the top electrode using false color. Elemental mappings for (c) were collected using EELS and (f) using EDS.

After fabrication, individual Josephson junction devices with a designed area of 3.1 μm$^2$ were contacted with Al wires and loaded into an adiabatic de-magnetization refrigerator (ADR) that has a base temperature of 0.25 K. Four-point-probe *I-V* data measured at 0.25 K across stacks containing thin Nb ($TaN_x$) seed and re-seed layers are shown in Figure 5a (Figure 5d), respectively. All devices demonstrate electrical tunneling behavior that is characteristic of unshunted *S-I-S* junctions—namely, we observe a hysteretic *I-V* behavior with a zero-voltage supercurrent branch, an Ohmic branch at high currents with a normal resistance ($R_N$), and a quasiparticle re-trapping branch that features a sharp current rise at the sum-gap voltage, $(\Delta_1(T) + \Delta_2(T))/e$. The sum-gap voltages we observe are consistent with a picture in which α-Ta provides the majority of the support for the superconducting wavefunctions in both the base electrode and top electrode regions of the stack (bulk single crystals of α-Ta show $2\Delta(T = 0)/e$ = 1.40 mV [42]). Sub-gap leakage was characterized by measuring the sub-gap resistance $R_{sg}$ at a voltage of 0.8 mV; for the Nb-seeded junctions studied in this work we observe an average $R_{sg}/R_N$ of 42.5, while the $TaN_x$-seeded junctions exhibit an average $R_{sg}/R_N$ of 15.2.

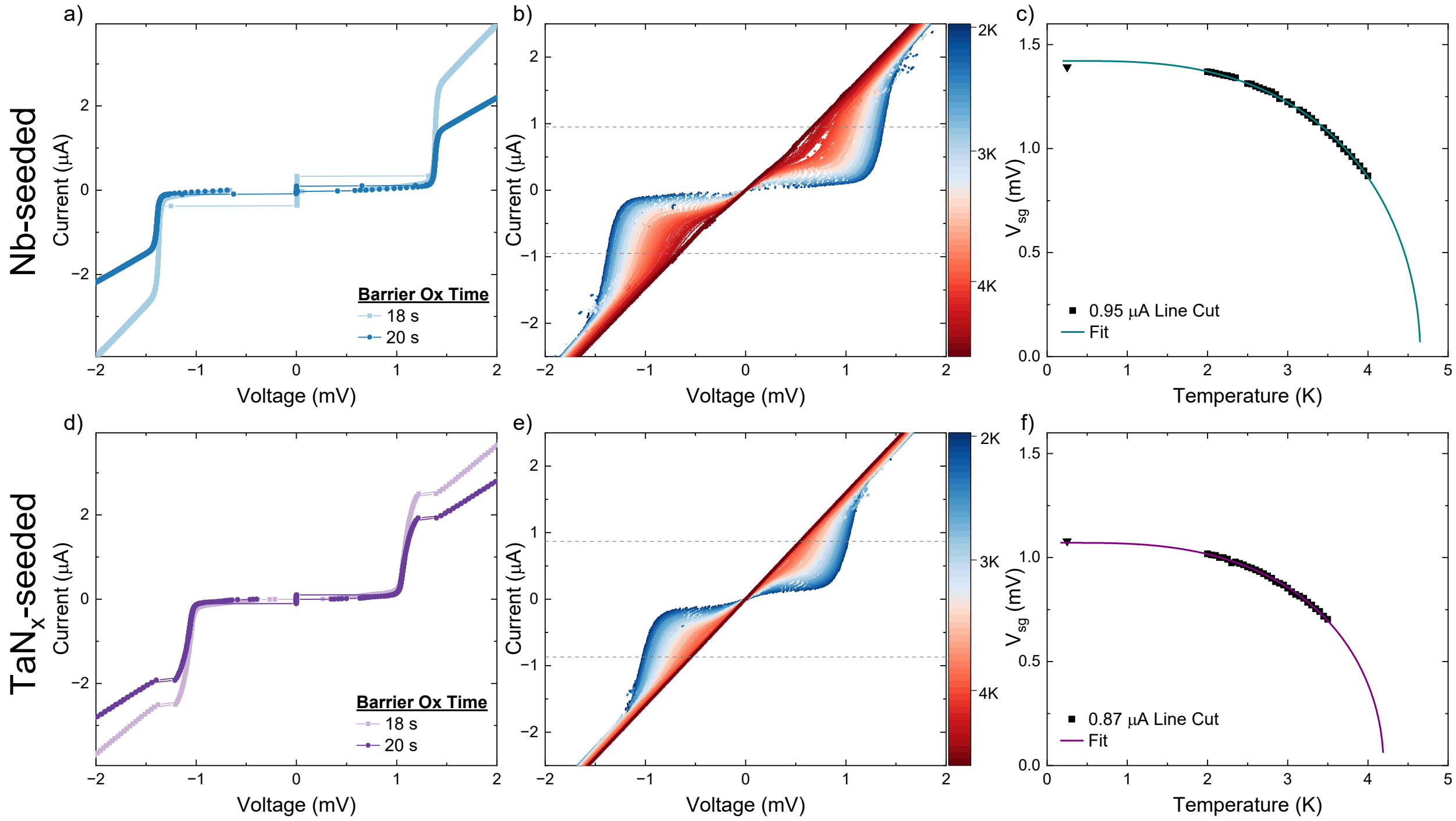

Figure 5 — Current-voltage characteristics of Nb and $TaN_x$ seeded junctions are shown in the top and bottom panel rows, respectively. (a, d) *I-V* curves measured at 0.25 K in an ADR. (b, e) Temperature dependent *I-V* curves collected between 2 K and 4.5 K in 0.05 K steps (cf. false color scales), measured in a separate 4He dry cryostat. (c, f) Sum-gap voltages $(\Delta_1(T) + \Delta_2(T))/e$ versus temperature extracted from the tunneling data shown in other panels. Solid lines are two-parameter fits to a BCS-like temperature dependence of the available data points (see main text), from base temperature up to approximately 90% of $T_c$.

To further investigate the tunneling behavior of these *S-I-S* junctions, illustrated in Figure 5(a, d), we performed a series of temperature-dependent *I-V* measurements, presented in Figure 5(b, e) for the stacks containing thin Nb- and $TaN_x$- seed layers, respectively. At $|V| > 1.5$ mV, the *I-V* characteristics exhibit linear Ohmic response at all temperatures. Below the superconducting $T_c$s of the two electrodes (4.0 – 4.5 K), the *I-V* characteristics begin to show non-linearities at low voltages, which develop into an abrupt rise in the quasiparticle current as the temperature is lowered further to 2 K, the base temperature of the PPMS cryostat used for these measurements shown in Figure 5(b, e). Specifically, at $T = 2$ K we find the voltage widths over which the current rises from 20% to 80% of its normal-state values $I_{ns}$ to be δV = 0.16 mV (0.21 mV) for the stacks containing Nb ($TaN_x$); at $T = 0.25$ K in Figure 5(a, d), these voltage widths become appreciably sharper, δV = 0.04 mV (0.10 mV). In all cases the normal-state values $I_{ns}$ used for thresholding are taken as the experimentally-measured current for the given data set at 1.5 mV and 1.2 mV for Nb- and $TaN_x$-seeded stacks, respectively.

In the case of conventional *s*-wave superconducting electrodes, the rise in the *S-I-S* tunneling current at the sum-gap voltage $(\Delta_1(T) + \Delta_2(T))/e$ is expected to be a discontinuous vertical jump at all temperatures $T < T_{c1}$, $T_{c2}$, because of singularities in the quasiparticle density of states for the superconducting electrodes immediately above their respective gap edges [43, 44]. In actual thin-film structures, however, this discontinuity is usually broadened into a finite-width feature versus voltage because of inhomogeneous superconducting properties across the device and quasiparticle lifetime broadening [21, 45]. To extract this superconducting sum-gap energy scale from the tunneling data, we find the intersection of each *I-V* characteristic acquired at a given temperature with the horizontal dashed lines in Figure 5(b, e), and plot the resulting points in Figure 5(c, f); these horizontal dashed lines are drawn at 50% of the respective normal-state $I_{ns}$ values, as defined above.

To extrapolate the measured sum-gap voltages to the saturated (very-low-temperature) regime, we fit the available $(\Delta_1(T) + \Delta_2(T))/e$ data to $2\Delta(T = 0)\tanh(1.74\sqrt{T_c/T - 1})$. This functional form approximates the full numerical solution of the self-consistent BCS gap equation to within 2% at all $T < T_c$ [46, 47, 48]. Note that we allow $\Delta(T = 0)$ to vary independently of $T_c$ in our fits, to account for the well-established empirical observation that $2\Delta_{\alpha\text{-Ta}}(T = 0)$ exceeds the universal weak-coupling mean-field (BCS) value of $3.53k_BT_c$ [46, 49] in both single crystals [42] and sputter-deposited thin films [23, 50] of α-Ta. We find that such two-parameter fits of the sum-gap voltages provide an excellent description of both the Nb and $TaN_x$ data sets, with inferred limiting values of $2\Delta(T = 0)/e$ = 1.42 mV (Nb) and 1.07 mV ($TaN_x$) that are consistent with the ADR *I-V* data shown previously in Figure 5(a, d). Note that this difference in sum-gap voltages between the Nb-seeded and $TaN_x$-seeded samples cannot be explained solely by the differences in $T_c$ noted earlier (Figure 1), as discussed in further detail in Supplementary Figure 4 [40]. The sum-gap voltage in Josephson junctions containing a thin superconducting layer close to the tunnel barrier is strongly influenced by the proximity effect, which depends on a number of different factors, including the thickness, coherence length, and critical temperature of the proximity layer, as well as the nature of its interface with the thicker superconducting electrode [51]. Consequently, it is not surprising to see a difference in sum-gap voltage between the Nb-seeded and $TaN_x$-seeded junctions, given the differences between Nb and $TaN_x$, and future investigations will focus on gaining a better understanding behind the mechanisms driving the sum-gap voltages in these junctions.

While junction measurements at cryogenic temperature are necessary for demonstrating the quality of the junctions, they are unable to provide the larger statistics necessary to evaluate the junction's manufacturability and scalability. We sample our process at the wafer scale by using the sum-gap voltages ascertained above to apply Ambegaokar-Baratoff (A-B)-style estimates of critical current densities from room-temperature resistance measurements as [52]: $I_c = J_cA = \pi(\Delta_1 + \Delta_2)/4eR_N$, where $A$ is the designed area of the junction, $R_N$ is the normal-state resistance and $(\Delta_1 + \Delta_2)/e$ is 1.42 mV for Nb-seeded junctions and 1.07 mV for $TaN_x$-seeded junctions as determined above. It should be noted that the A-B formula assumes weak-coupling superconductors, and deviations from this assumption result in multiplicative scale factors that reduce the maximum $I_c$ relative to the values calculated from this approximation [53, 54]. Additionally, the $R_N$ for this equation is typically taken as the resistance of the junction with the superconducting electrodes in their voltage-driven normal states at temperatures below $T_c$, thus, a correction factor must also be applied to room-temperature measurements, which has been established experimentally below.

To explore a broader process space, we fabricated multiple wafers with varied plasma oxidation time. Each wafer contained approximately 28,500 individual junctions, covering six distinct design areas ranging from 0.4 μm$^2$ to 3.1 μm$^2$, for which $R_N$(295 K) values were measured. Comparing the normal-state resistances measured at 0.25 K for select devices to the room temperature resistance measured, a residual resistivity ratio for the $TaO_x$ barrier of $R_N$(295 K)/$R_N$ = 0.4 is observed across growth conditions. Combining this empirical ratio with the A-B formula from above gives $I_c = \pi(\Delta_1 + \Delta_2) * 0.4/4eR_N(295\ K)$, which is plotted as the vertical axis of the data points in both panels of Figure 6. The summary of the fitting results from Figure 6, as well as the same analysis performed on wafers with lower oxidation times, are aggregated in Figure *7*.

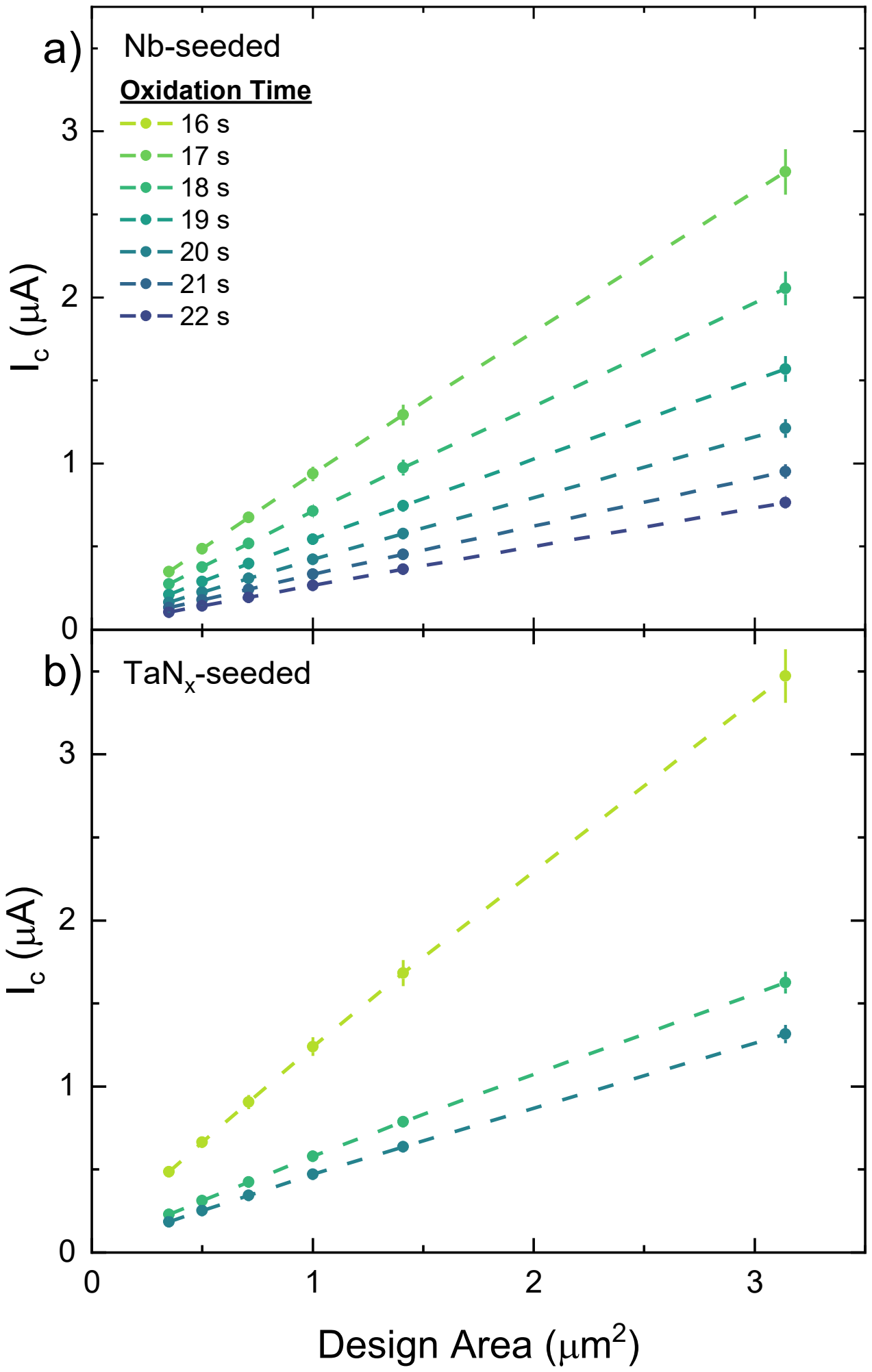

Figure 6 —Estimated $I_c$ values calculated from room temperature $R_N$ data using the Ambegaokar-Baratoff formula for six different sizes of junctions patterned from both (a) Nb-seeded Ta and (b) $TaN_x$-seeded Ta stacks, respectively. Dotted lines are linear fits of the data; the slopes of which provide the $J_c$ values for the given stack and plasma oxidation time (false color scale).

Figure 6 shows that 1/$R_N$(295 K) scales proportionally with the nominal design area of the junction, irrespective of the plasma oxidation time and choice of re-seed layer, suggesting $TaO_x$ barrier layers that are locally homogeneous across lateral in-plane areas ranging from 0.4 μm$^2$ to 3.1 μm$^2$. Statistics on individual reticles of 20 mm x 20 mm lateral in-plane scale were focused on the smallest junctions (designed area of 0.4 μm$^2$). Electrical opens or shorts cause extreme outliers in the dataset which are not representative of the actual junction resistance, but rather are resultant from wafer-scale test configuration issues, such as poor probe touchdown contact. To provide a more accurate representation of true junction process spread in the presence of these outliers, the median absolute deviation of the entire population on each 5 mm x 5 mm chip was used to estimate the standard deviation of a normally distributed population of yielding junctions. The most central chip on each wafer had the lowest variation, achieving standard deviations within the range of 2.0% - 3.9% in the distributions of $1/R_N$(295 K) for all wafers measured; no significant variation in spread was observed between the two seed layers used. Taken together, these measurements demonstrate the uniformity of the $TaO_x$ barriers produced by our plasma oxidation process across in-plane length scales relevant for an industrial foundry, as well as the reliability of the device patterning processes used to define the in-plane areas of the Josephson junctions.

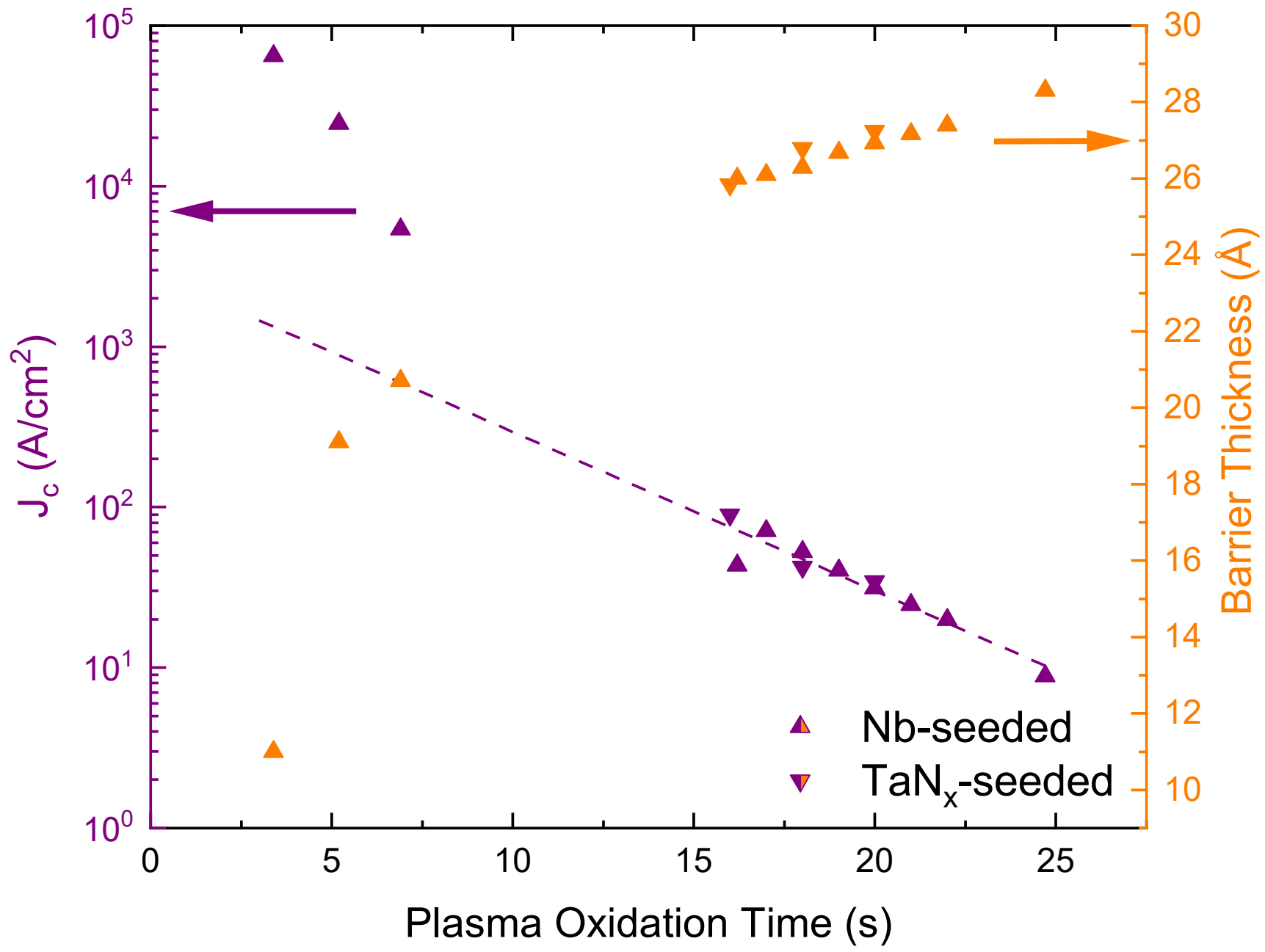

Figure 7 — Measured $J_c$ (left, purple) and $TaO_x$ barrier thickness, $t$, (right, orange), based on XRR modeling, for each of the samples shown. The dotted line is a log-linear fit of the $J_c$ for oxidation times ≥ 16 s.

Having established the reliability of the process, we now examine the range of achievable $J_c$ [21]. In Figure *7*, we show our calculated $J_c$ as a function of plasma oxidation time (i.e. barrier thickness). The $J_c$ of the Nb-seeded junctions ranged from 64,000 A/cm$^2$, at our shortest oxidation time (3.4 s), to 10 A/cm$^2$ when oxidized for 24.7 s. A much smaller range was explored for the $TaN_x$-seeded junctions, demonstrating a critical current density of 90 A/cm$^2$ at 16 s of plasma oxidation to 34 A/cm$^2$ with a 20 s oxidation. We find that the barrier thickness is in a linear growth regime for plasma oxidations ≥ 16 s, and observe good agreement between the $TaN_x$-seeded and the Nb-seeded JJs in this regime. As such, this provides a well-controlled process where $J_c$ can be tuned over an order of magnitude, thus providing a broad design space for Josephson junction-based electronics.

In general, superconducting qubit architectures require junctions with $I_c$ < 100 nA while maintaining a junction cross-sectional area of less than 1 μm$^2$ to limit the volume of the barrier. We have demonstrated this feasibility by measuring an $I_c$ of 40 nA with a junction area of 0.4 μm$^2$ ($J_c$ of 10 A/cm$^2$). Moreover, given the longest plasma oxidation time grown for this work (24.7 s) remains in the linear barrier growth rate regime while demonstrating corresponding continued log-linear reduction in $J_c$, it is reasonable to suggest further reduction in $J_c$ is possible with longer oxidation times should that be required for a given application.

## III. Conclusions

In this work, we have presented *S-I-S* Josephson junctions with α-Ta electrodes and a $TaO_x$ barrier. We grew base and top α-Ta electrodes using seed layer engineering, showing both Nb and $TaN_x$ are viable seed layer choices, and particularly emphasize that at room temperature a re-seed layer is required to nucleate α-Ta for the top electrode. However, we note that the choice of seed layer can have subtle implications on the electrical tunneling behavior of the junctions. We used room-temperature probing of variable junction areas, combined with Ambegaokar-Baratoff physics, to characterize the wafer scale critical current densities. For 0.4 $\mu m^2$ junctions, we found junction spreads as low as 2.0% across the 20 x 20 mm chip relevant for device performance. The entire process was carried out using standard industry foundry processing using 150 mm wafer tooling, which is an important step towards proving that Ta-based junctions can be implemented at scale. We have successfully fabricated and tested junctions with α-Ta electrodes and a $TaO_x$ insulating tunnel barrier, demonstrating that this is a material platform which merits further investigation and studies for potential application in classical and quantum superconducting technologies.

## Acknowledgements

We thank Justin Hackley, Alex Krick, Monica Lilly, Christopher Pinion, Christopher Rotella, Patrick Warner, and Daryna Soloviova for invaluable discussion over the course of this work. This work was supported in part by the United States Government.

## Author List

Trent Borman, Cameron Bullock, Cole Chason, Bradley Christensen, Jonathan Cochran, Emad Din, Evan Donohue, Jason Flagg, A. L. Graninger, Eric Jones, Shawn Keebaugh, Erin Kennedy, Robert Klaes, John Kula, Jason Lapano, Melissa Loving, Christopher Pasco, Devika Pathak, Tam Pham, Timothy Pillsbury, Robert Pownall, Paige Quarterman, Jacob Ruf, Jade Simmons, Nathan Siwak, Joel Strand, Yong Mo Yang

Contact author: paige.quarterman@ngc.com

## Methods

### Film Growth

Prior to the deposition of metal and insulating layers in the junction stacks, Si(100) wafers were pretreated with an *ex-situ* chemical clean in diluted hydrofluoric acid. Ta and Nb layers in the stack were deposited from single targets (>99.99% purity) in a CANON-ANELVA magnetron sputter system having a base pressure < 5E-09 Torr, using Ar process gas, onto substrates water cooled to 25 °C. The film stacks were deposited at a low working pressure (< 5E-04 Torr). Where applicable, $TaN_x$ layers were deposited using a reactive sputter deposition process in the same tool as above, with a mixture of Ar and $N_2$ gases. The gas flow rate ratio, $R = Ar/(Ar + N_2)$, is the primary process parameter that determines the stoichiometry ($x$) and crystallinity of the $TaN_x$ layer [35, 36]. For all films and devices shown in this paper we chose $R = 0.4$. This $R$ value results in a $TaN_x$ layer with $0.5 < x < 1$ that terminates at sharply-defined bounding interfaces (≈ 5 Å - 10 Å) and consists of a highly-disordered stacking sequence of Ta and N at the level of individual atomic planes (characteristic room-temperature resistivities of 500 μΩ-cm - 1000 μΩ-cm).

These features are consistent with numerous prior literature reports on similarly-prepared $TaN_x$ films [36, 55]. Following base electrode deposition, the insulating $TaO_x$ tunnel barrier in the *S-I-S* stack was created using an Ar + $O_2$ plasma oxidation process in a separate chamber in the ANELVA tool.

**X-ray Scattering**

X-ray diffraction and reflectometry measurements were collected using monochromatic Cu-$K_{\alpha 1}$ radiation ($\lambda$ = 1.54059 Å), using Panalytical X'pert MRD XL and Bede lab-source diffractometers. Reflectometry data were modeled using the REFL1D software package to extract film thicknesses and interfacial roughnesses. Coupled 2Θ-ω scans ($\omega = 2\theta/2$), shown in Figure 1 of the main text and Supplementary Figure 1, were acquired with the momentum transfer aligned along the specular crystal truncation rods defined by the Si(100) substrates. 2Θ-ω scans displayed in Figure 2 of the main text were acquired with a small ω offset for all samples, $\omega = 2\theta/2 - 0.5°$, to eliminate Si multiple scattering peaks observed at $2\theta = 32.97°$ in Figure 1, while having minimal impact to Ta peak intensities, due to their broad ω rocking curves [40].

**Josephson Junction Fabrication**

We fabricated Josephson junctions by first patterning the Ta multilayer stacks with DUV (248 nm) photolithography and dry etching pillars into the stack with $BCl_3/Cl_2$ etch chemistry. The pillar etch was stopped once the etch broke through the multilayer stack and exposed the base α-Ta layer. We stopped the etch at this point so that the base α-Ta layer could be utilized as a routing layer to electrically access the base electrode of the junction. We then used photolithography to pattern the base α-Ta layer and etched down to the high-resistance Si substrate using the same dry etch conditions as done for the pillar etch. We grew silicon-oxide using plasma-enhanced chemical vapor deposition for electrical isolation between the junction and routing layers. Next, we patterned vias and etched through the silicon oxide and to the top of the junction electrode and base α-Ta routing layer. Once patterned, we sputtered a thick Nb film to provide sufficient overfill of the vias and so that it could be used for additional routing out to electrical contact pads. This Nb film was etched and formed isolated contacts between to the top of the Ta junction and the bottom routing layer. Finally, we patterned Al pads using typical photolithography, evaporation, and lift-off processing to complete the device fabrication. Post fabrication of the 150 mm wafers, each of the twenty-six 20 x 20 mm reticle was diced into sixteen 5 x 5 mm chips.

**Electron Microscopy**

Tilted SEM images were carried out in a ThermoFisher Helios 5 UX PFIB system using a 2 kV source. All STEM samples were prepared using a Ga-based focused ion beam and standard lift-out procedures. STEM HAADF imaging, EDS, and EELS were carried out using a 200 keV spherical-aberration corrected source. EDS mapping was performed using as SuperX system, and EELS maps were collected with a Gatan Enfinium ER spectrometer capturing energy loss windows for the zero-loss region, oxygen edge (approximately 170 to 2200 eV), and the niobium and tantalum edges (approximately 1000 eV to 3070 eV). EELS was preferred in our elemental analysis of the Nb-seeded sample due to a need to disambiguate X-ray signals (Si K and Ta M peaks) from surrounding Si in other portions of sample (not shown in this work). The focus of our analysis on the $TaN_x$-seeded sample was solely at the interface of the oxide barrier, diffusion barrier, and re-seed layer allowing for EDS to be performed without worry of Si peak overlap.

**Electrical Transport Characterization**

We measured cryogenic transport across patterned *S-I-S* Josephson junction stacks in an adiabatic demagnetization refrigerator at fixed $T$ = 0.25 K using a four-terminal wiring configuration. Current is supplied via a Keysight B2962B® Power Source and voltage is measured via a Keysight 34461A® Digital Multimeter. Low-pass filters with an designed 26 Hz cutoff were applied to both the current output and the voltage input to eliminate AC noise.

Additional cryogenic transport measurements were performed at higher variable temperatures ($T$ = 2 - 300 K) using a Quantum Design Physical Properties Measurement System (PPMS®) model 6000. Four-point resistance vs. temperature measurements on un-patterned films were performed using the DC resistivity option. Temperature-dependent *I-V* measurements across patterned *S-I-S* JJ stacks were performed using the electrical transport option (ETO), which utilizes a digital lock-in technique.

High-throughput room-temperature probing of patterned *S-I-S* JJ stacks was accomplished using a Keithley S530® Parametric Test System equipped with a Keithley Model 2636B® 2-channel source meter unit and a Keithley Model 7510® digital multimeter for readout. The unit uses a Keithley Model 707B® switch matrix to enable the fast readout of multiple junctions.

**Magnetometry**

Meissner field expulsion measurements were collected in a Quantum Design MPMS®3 SQUID magnetometer. Measurements were taken by warming from the base temperature of 2 K, through the superconducting transition temperature, up to 6 K with an applied field of 5 Oe after cooling in zero field. The same 5 Oe field was left on and the sample was then field cooled back to 2 K.

# SUPPLEMENTARY INFORMATION

## I. XRR and XRD

Supplementary Figure 1 shows additional x-ray diffraction (XRD) data for the samples characterized in Figures 1 and 2 of the main text. Scattering from the α-Ta-220 Bragg reflection is maximized at $2\Theta_{peak}$ values indicated by vertical dashed lines in panels (a, d) for the Nb- and $TaN_x$-seeded stacks, respectively. Comparing these measured $2\Theta_{peak}$ values with the expected value of 82.5° for bulk single crystals of α-Ta demonstrates that the α-Ta films in our samples are subject to high out-of-plane tensile strains, on the order of +1% relative to bulk; these strains result from large biaxial in-plane compressive stresses induced by sputter depositing Ta at low working pressures of Ar process gas (see Methods). Specifically, we measure $2\Theta_{peak}$ values of 81.5° (81.4°) for the Nb-seeded ($TaN_x$-seeded) base-electrode-only samples, and 81.7° (81.55°) for the two Nb-seeded ($TaN_x$-seeded) *S-I-S* stacks that include a $TaO_x$ layer formed by 18 seconds of Ar + $O_2$ plasma oxidation; these values correspond to average out-of-plane strains of +1.0% (+1.1%) for the Nb-seeded ($TaN_x$-seeded) base-electrode-only samples, and +0.80% (+0.96%) for the full *S-I-S* stacks including the effects of plasma oxidation. Furthermore, we note that the α-Ta-220 peak shapes versus 2Θ show broad distributions of inhomogeneous microstrain that are asymmetrically skewed towards larger (i.e., more bulk-like) angles; the full width at half maximum (FWHM) values of the peaks for all samples shown here are about $\delta(2\Theta)$ = 1.0°.

To probe the mosaic spread of the α-Ta crystallites, we integrate the scattering over a range of 2Θ values spanning $2\Theta_{peak} \pm 0.5°$ (cf. the gray-shaded rectangles in Supplementary Figure 1 panels (a, d)) and vary the angle of the incident x-ray beam (ω) over a ± 15° range relative to the surface normal ($\omega_{CTR} = 2\Theta_{peak}/2$) defined by the Si(100) substrates. The resulting ω rocking curves are plotted in panels (b, e) of Supplementary Figure 1 for the Nb- and $TaN_x$-seeded stacks, respectively, and the ω FWHM values of the peaks are recorded in the insets of (b, e). The α-Ta crystallites in the Nb-seeded stacks consistently exhibit 2x - 2.5x sharper peaks versus ω (i.e., significantly reduced mosaic spread) relative to the α-Ta crystallites in their $TaN_x$-seeded counterparts.

Finally, to probe how effectively the 30 Å-thick Nb and $TaN_x$ re-seed layers (above the amorphous $TaO_x$ layer) stabilize well-oriented α-Ta crystallites in the 1 kÅ-thick Ta layers above, we compute the integrated intensities over the angular ranges $\{2\Theta = 2\Theta_{peak} \pm 0.5°, \omega = \omega_{CTR} \pm 15°\}$ for the x-ray scattering data shown in (a-b, d-e), and plot the results in panels (c, f) of Supplementary Figure 1. In these plots, an integrated intensity approaching a value of ‘2’ would correspond to a preferentially *HH0*-out-of-plane-oriented α-Ta top electrode that exactly replicates the distribution of grain tilts observed for the as-deposited base electrode. By contrast, an integrated intensity near a value of ‘1’ corresponds to a top electrode that either (1.) does not contain a significant volume fraction of α-Ta, or (2.) consists of highly polycrystalline (i.e., very poorly-oriented) crystallites of α-Ta; in other words, ‘1’ indicates the top electrode does not meaningfully scatter x-rays into the α-Ta-220 Bragg reflection, such that all of the measured contributions to that peak’s integrated intensity can be accounted for by x-rays scattering off the base electrode. We find α-Ta-220 integrated intensities of 1.7 (1.4) for the full *S-I-S* stacks containing Nb ($TaN_x$) re-seed layers, and α-Ta-220 integrated intensities of 1.1 (1.0) for the corresponding *S-I-S* stacks without Nb ($TaN_x$) re-seed layers.

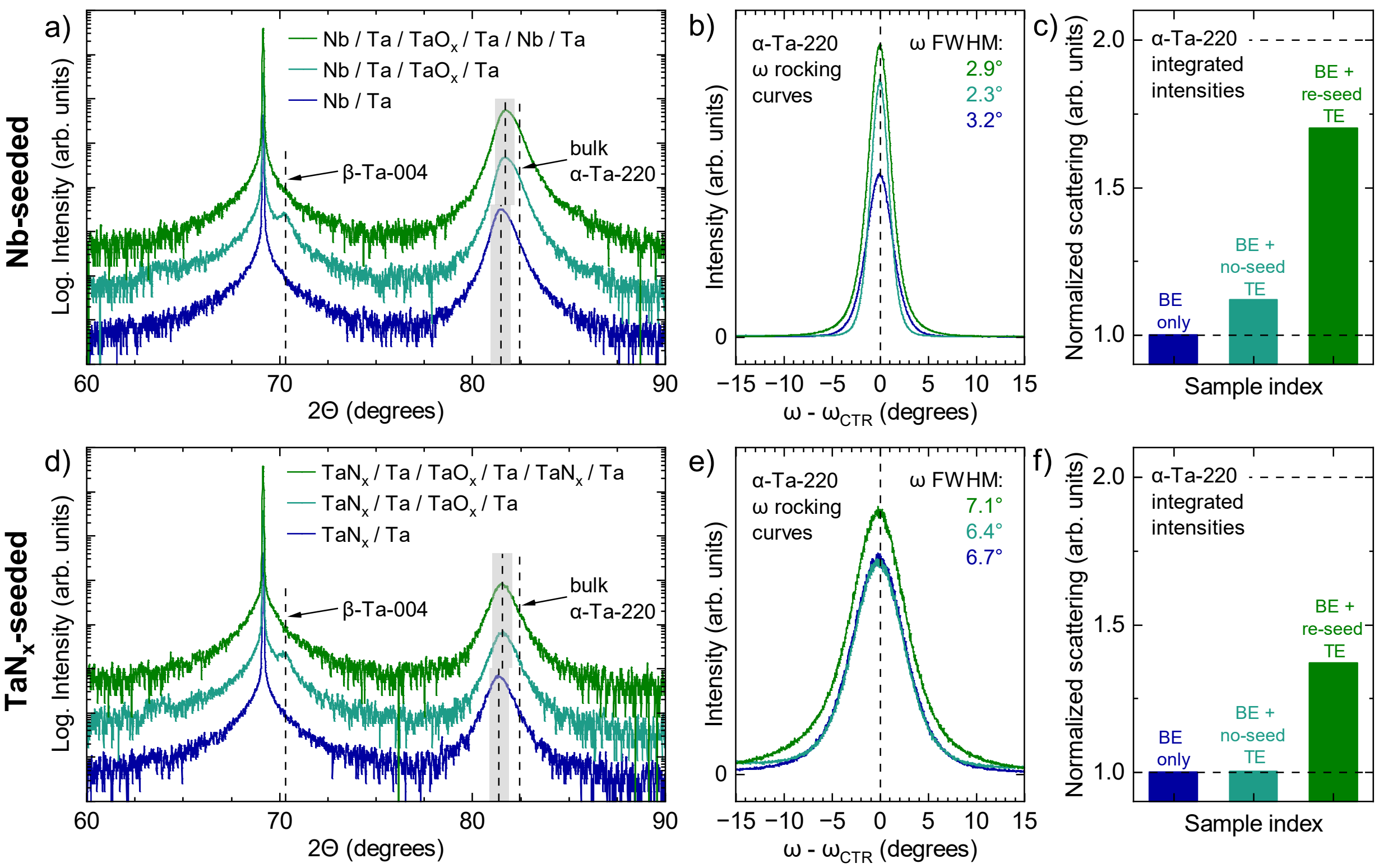

Supplementary Figure 1 – (a) Coupled 2Θ-ω XRD scans (ω = 2Θ/2) for the Nb-seeded stacks indicated in the legend, vertically offset for clarity and plotted versus logarithmic vertical scales. (b) ω rocking curves centered on the α-Ta-220 peaks for the samples in (a), plotted versus a linear vertical scale with no offsets between samples. (c) Integrated intensities (areas under the curves) for the data shown in (b), with all three integrated intensity values uniformly multiplied by a scale factor that makes the base-electrode-only sample equal to 1. Data in panels (d – f) are the analogous measurements and analysis from (a – c), now for $TaN_x$-seeded stacks. Note that the vertical scale of the ω rocking curves in (b) is six times taller than the vertical scale in (e), and the integrated intensity of the α-Ta-220 peak for the Nb-seeded base-electrode-only sample in (c) is 2.1 times larger than its $TaN_x$-seeded counterpart in (f).

We used XRR to map the plasma oxidation times of Ta to a corresponding $TaO_x$ thickness. We used a simplified thin film stack for this calibration, which was nominally: Si(substrate) / Ru(25) / Ta(100) / $TaO_x$(*t*) / Ru(25), where the thicknesses are in Angstroms, and the oxide thickness, *t*, was varied by increasing the plasma oxidation time from 3.4 s to 24.7 s. We model-fit the XRR datasets simultaneously by requiring the scattering length density (SLD) for each layer to be shared across the full sample set. We found the data can be well modelled by varying the SLD no more than ±5% of the bulk calculated SLD, except for $TaO_x$ where the large variation in SLD range is an approximation of the stoichiometry. We find a Ta SLD value of 106 x $10^{-6}$ $Å^{-2}$, which is near the bulk value, and we find a fit SLD of $TaO_x$ to be 57 x $10^{-6}$ $Å^{-2}$, where the reduced SLD is a consequence of oxygen inclusion into the film. In an alternative model, we fixed the SLD to the bulk calculated values and found that the data cannot be sufficiently fit to key features. The imaginary component of the scattering length density (iSLD) was forced to vary by the same scale factor as the real component, that is not fit individually from the real component. The individual layer interfacial roughnesses and film thicknesses were allowed to independently vary to account for both small variations between growth conditions across samples, and subtle variations in instrument repeatability in terms of resolution and wafer curvature. The data and corresponding fits are shown in Supplementary Figure *2*. A summary table of the parameters extracted is provided in Supplementary Table 1. We note that the interfacial roughnesses of the stacks used in the main text are expected to be larger given the increased film thicknesses of the Ta layers.

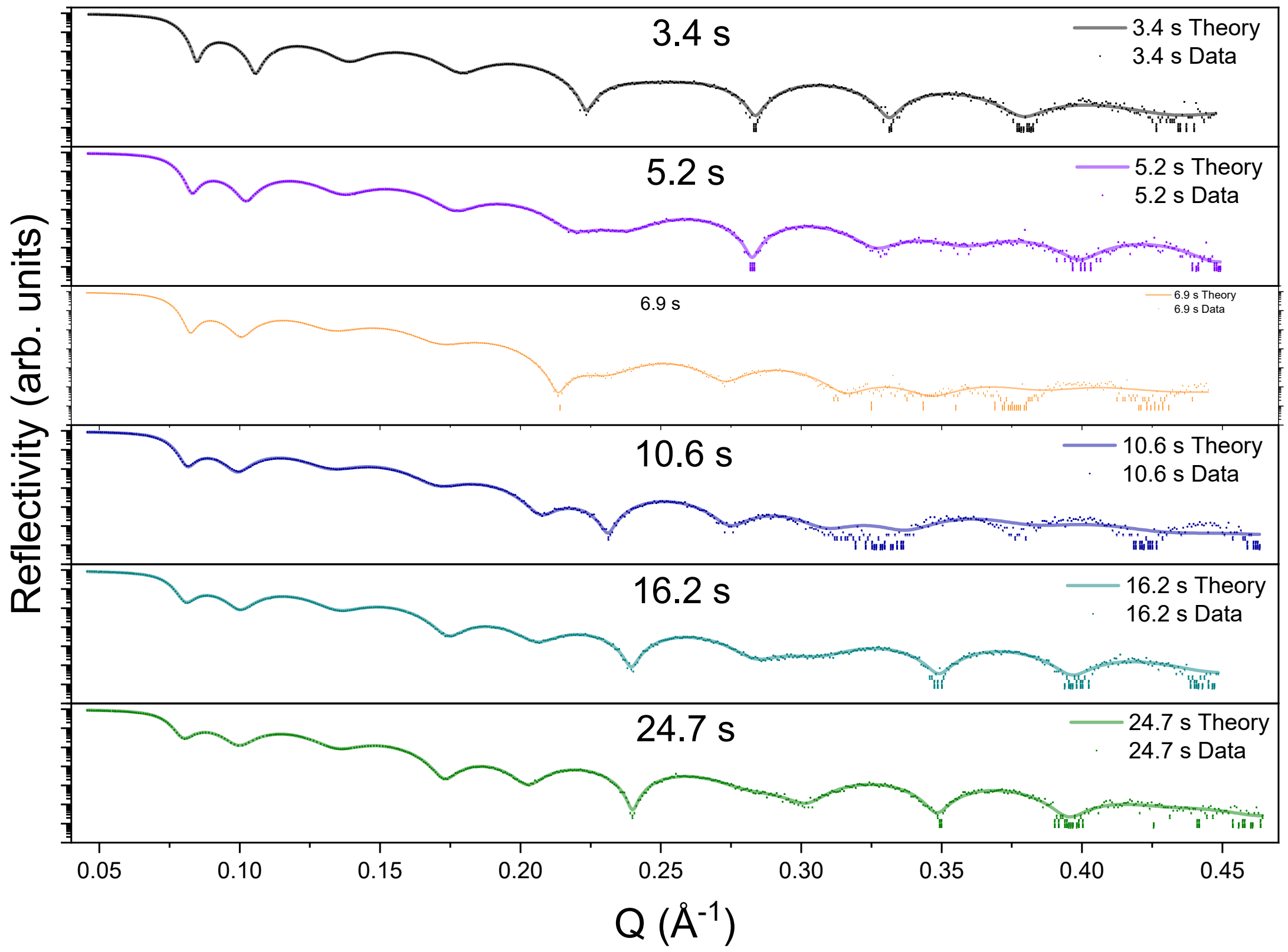

Supplementary Figure 2 – XRR for representative blanket films of Si/Ru/Ta/$TaO_x$/Ru stacks that were used to calibrate the $TaO_x$ thickness as a function of plasma oxidation time. Points show XRR data collected and the overlaid line shows the model-fit corresponding to parameters in Supplementary Table 1.

| | **Si Substrate** | | **Bottom Ru Layer** | | | **Ta Layer** | | | **$TaO_x$ Layer** | | | **Top Ru Layer** | | |
|---|---|---|---|---|---|---|---|---|---|---|---|---|---|---|
| Oxidation time (s) | Interface (Å) | SLD ($10^{-6}$ Å$^{-2}$) | Interface (Å) | SLD ($10^{-6}$ Å$^{-2}$) | t (Å) | Interface (Å) | SLD ($10^{-6}$ Å$^{-2}$) | t (Å) | Interface (Å) | SLD ($10^{-6}$ Å$^{-2}$) | t (Å) | Interface (Å) | SLD ($10^{-6}$ Å$^{-2}$) | t (Å) |
| 24.7 | 3 | 20 | 7 | 90 | 23 | 3 | 107 | 88 | 4 | 57 | 28 | 6 | 93 | 25 |
| 16.2 | 3 | 20 | 7 | 90 | 22 | 3 | 107 | 89 | 4 | 57 | 26 | 6 | 93 | 25 |
| 10.6 | 6 | 20 | 8 | 90 | 24 | 4 | 107 | 94 | 4 | 57 | 23 | 6 | 93 | 26 |
| 6.9 | 6 | 20 | 7 | 90 | 24 | 4 | 107 | 94 | 3 | 57 | 20 | 6 | 93 | 26 |
| 5.2 | 3 | 20 | 7 | 90 | 23 | 4 | 107 | 91 | 4 | 57 | 19 | 6 | 93 | 26 |
| 3.4 | 3 | 20 | 7 | 90 | 23 | 4 | 107 | 93 | 5 | 57 | 12 | 6 | 93 | 27 |

Supplementary Table 1 – Extracted parameters based on model-fitting the data shown in Supplementary Figure 2 with samples grown using varied plasma oxidation time. The SLD and iSLD values were fixed across samples.

## II. Electron Microscopy

Scanning transmission electron microscopy (STEM) was used to verify the uniformity and integrity of the various junction barriers and seed layers of a representative device. High angle annular dark field images (HAADF) show continuous and well-defined $TaO_x$ barriers, despite the roughness that is apparent at lower magnifications, see Supplementary Figure 3 panels a) and d). This roughness also contributes to the seemingly higher-Z contrast in the $TaO_x$ barrier in a) vs d). Comparison with other portions of the sample, however, show comparable contrasts between the two samples. Electron energy loss spectrometry (EELS) was preferred in our analysis of the Nb-seeded sample due to the need to disambiguate X-ray signals (Si K and Ta M peaks) from surrounding Si in other regions of the sample that were imaged (not shown). The focus of our analysis on the $TaN_x$-seeded sample was solely at the interface, allowing for energy dispersive X-ray spectrometry (EDS) to be performed without worry of Si peak overlap. Both EELS and EDS elemental mappings (Supplementary Figure 3 panels b and e) and their corresponding integrated line-scans through the layers (Supplementary Figure 3 panels c and f) have some level of beam broadening and surface roughness effects, as discussed above, which make accurate estimation of the oxygen content at the barrier difficult.

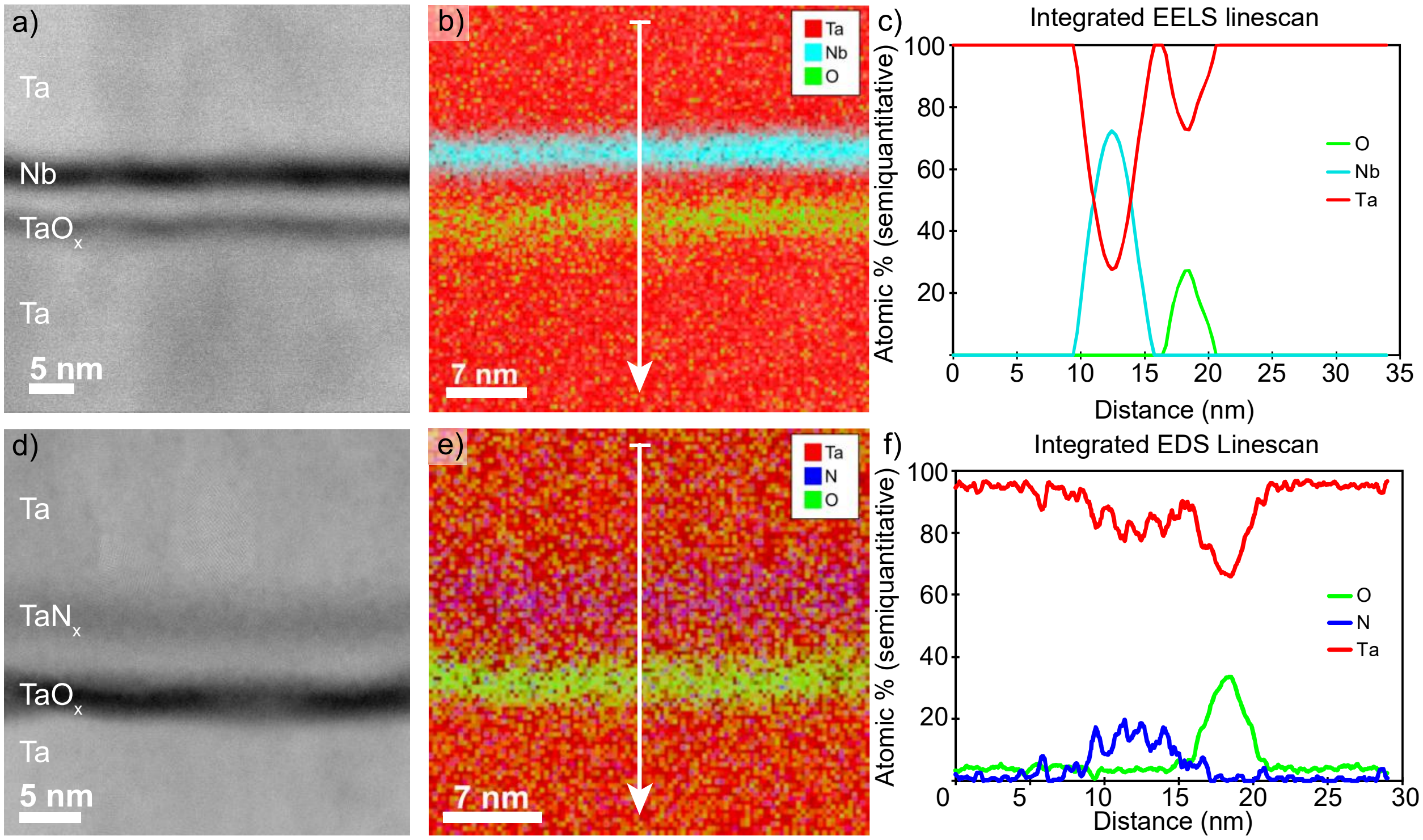

Supplementary Figure 3 – HAADF STEM of the $TaO_x$, and re-seed layer for (a) Nb and (d) $TaN_x$. EDS map for the corresponding HAADF image (b) and (e), which are same as from Fig 4. In the main text, to provide context for the depth profile EDS line scans. (c,f) Integrated line scans as a function of depth for O, Nb, and Ta (0 nm references the top of the EDS map from (b,e), for the two re-seed cases of Nb and $TaN_x$ seeding).

## III. Additional Magnetic and Electrical Characterization

Supplementary Figure 4 shows the same data from the temperature-dependent current-voltage curves from Figure 5 of the main text superimposed on the same plot along with the fit of the Nb-seeded sample scaled by the ratio of $T_c$s of the two material stacks. Naïvely, given the BCS theory relationship between $T_c$ and energy gap ($\Delta(0) = 1.764k_BT_c$) one might expect the variation in $T_c$ to correspondingly explain the variation in sum-gap voltage, but the data in Supplementary Figure 4 show that this explanation is not sufficient to account for the magnitude of the discrepancy between the datasets.

The upper critical field, $H_{c2\perp}$, of the materials for each deposition method was probed by measuring moment vs. field on Nb- and $TaN_x$-seeded Ta blanket films in a range of temperatures between 2 K and 4.5 K with the field applied perpendicular to the sample plane. Supplementary Supplementary Figure 5a) shows a subset of these measurements performed on $TaN_x$-seeded Ta samples, from which the upper critical field at each temperature can be extracted. Supplementary Figure 5b) shows these extracted $H_{c2}$ plotted against temperature. A linear fit to these data gives the zero-temperature out-of-plane upper critical field, $H_{c2}(0)$. From this, an in-plane Ginsburg-Landau coherence length can be found from $\mu_0H_{c2}(0\ \mathrm{K}) = \Phi_0/(2\pi\xi^2_{GL})$. This methodology gives values for $H_{c2}(0)$ of 1.19 T and 0.91 T, and values for $\xi_{GL}$ of 16.6 nm and 19.0 nm, for the $TaN_x$- and Nb-seeded samples, respectively.

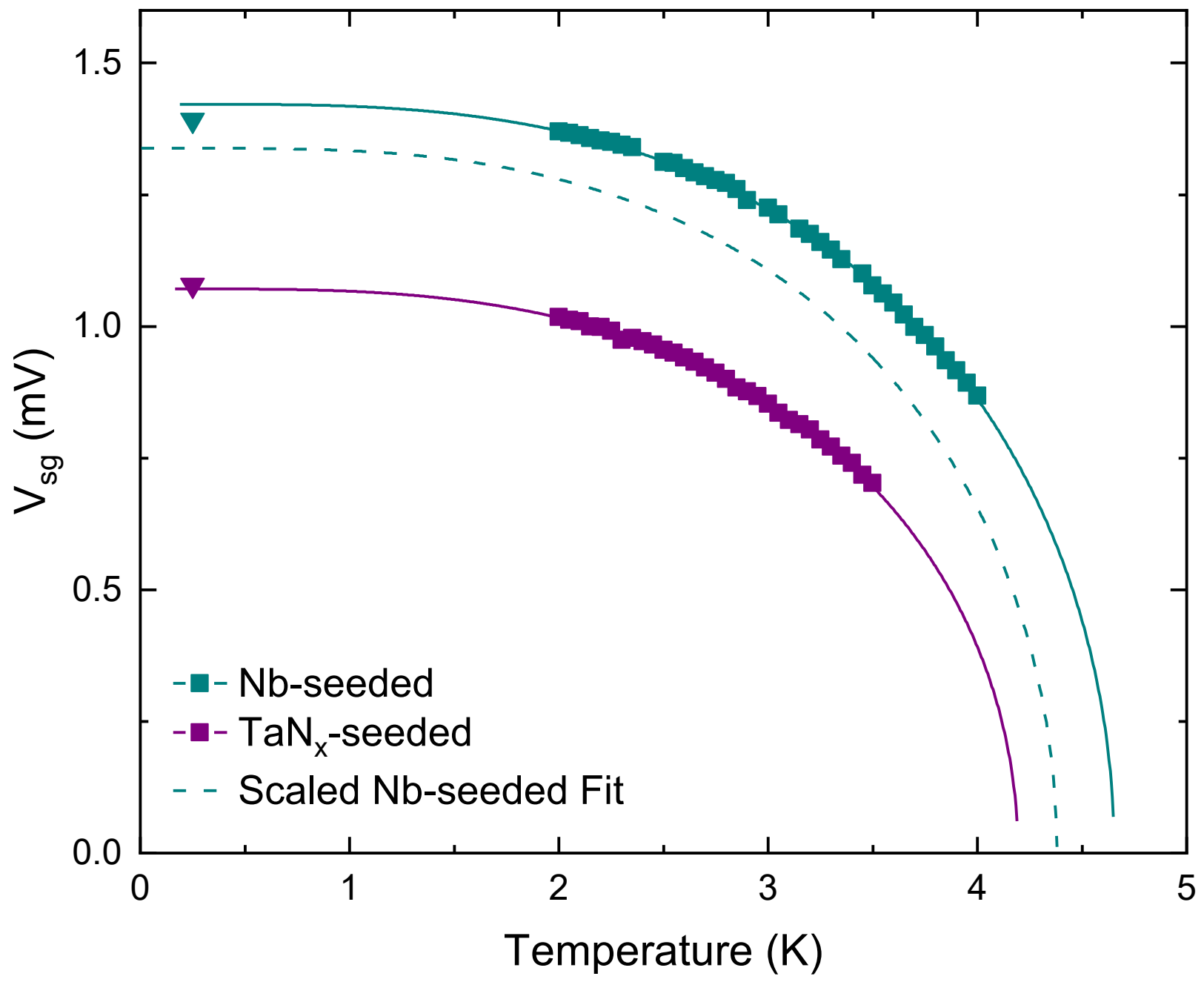

Supplementary Figure 4 — Data from main text Figure 5 c) and f) replotted together, showing difference in sum gap voltage between Nb-seeded (teal) and $TaN_x$-seeded (purple) samples. The dashed green line is a rescaling of the Nb-seeded fit by the ratio of the critical temperatures of the sheet films, i.e. $T_c(TaN_x\text{-seed})/T_c(\text{Nb-seed})$.

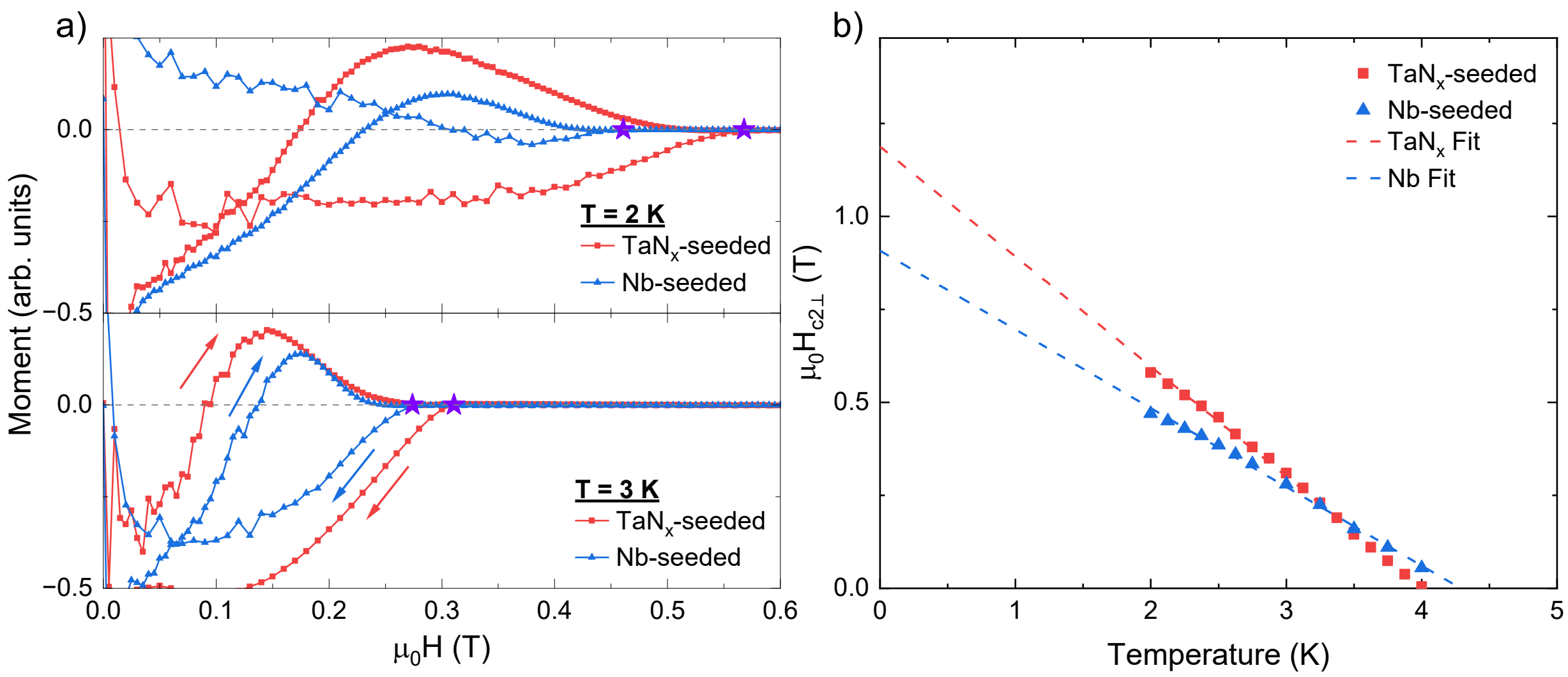

Supplementary Figure 5 — Moment vs field measurements for $TaN_x$- and Nb-seeded Ta blanket films were performed at temperatures between 2 K and 4.5K with the field applied perpendicular to the sample plane, data from 2 K and 3 K in (a). Defining downsweep onset $H_c$ at the point of first deviation greater than 3x the noise, and $H_{c2}$ as the previous datapoint (as indicated by the purple stars), the $H_{c2}$ data from all temperatures is plotted in (b). A linear fit was performed to extrapolate the upper critical field at zero temperature, $H_{c2}(0)$.